\documentclass[iop]{emulateapj}

\usepackage{graphicx}
\usepackage{natbib}

\def\lsim{\mathrel{\hbox{\rlap{\lower.55ex \hbox {$\sim$}}\kern-.0em
\raise.4ex \hbox{$<$}}}} 
\def\gsim{\mathrel{\hbox{\rlap{\lower.55ex \hbox {$\sim$}}\kern-.0em
\raise.4ex \hbox{$>$}}}}

\def\l{$\lambda$}

\shorttitle{NOT polarimetry of SN 2015bn}
\shortauthors{Leloudas et al.}

\begin{document}

\title{Time-resolved polarimetry of the superluminous  SN 2015bn with the Nordic Optical Telescope}

\author{
Giorgos Leloudas\altaffilmark{1,2}, 
Justyn~R.~Maund\altaffilmark{3},
Avishay Gal-Yam\altaffilmark{1},
Tapio Pursimo\altaffilmark{4},
Eric Hsiao\altaffilmark{5},
Daniele Malesani\altaffilmark{2,6},
Ferdinando Patat\altaffilmark{7},
Antonio de Ugarte Postigo\altaffilmark{8,2},
Jesper Sollerman\altaffilmark{9},
Maximilian~D.~Stritzinger\altaffilmark{10},
and J.~Craig Wheeler\altaffilmark{11}
}

\altaffiltext{1}{Department of Particle Physics and Astrophysics, Weizmann Institute of Science, Rehovot 7610001, Israel}
\altaffiltext{2}{Dark Cosmology Centre, Niels Bohr Institute, University of Copenhagen, Juliane Maries vej 30, 2100 Copenhagen, Denmark}
\altaffiltext{3}{The Department of Physics and Astronomy, University of Sheffield, Hicks Building, Hounsfield Road, Sheffield, S3 7RH, UK}
\altaffiltext{4}{Nordic Optical Telescope, Apartado 474, E-38700 Santa Cruz de La Palma, Santa Cruz de Tenerife, Spain}
\altaffiltext{5}{Department of Physics, Florida State University, Tallahassee, FL 32306, USA}
\altaffiltext{6}{DTU Space, National Space Institute, Technical University of Denmark, Elektrovej 327, 2800 Lyngby, Denmark}
\altaffiltext{6}{European Southern Observatory, Karl-Schwarzschild-Strasse 2, 85748 Garching, Germany}
\altaffiltext{7}{Instituto de Astrof\' isica de Andaluc\' ia (IAA-CSIC), Glorieta de la Astronom\' ia s/n, E-18008, Granada, Spain}
\altaffiltext{8}{The Oskar Klein Centre, Department of Astronomy, Stockholm University, AlbaNova, 10691 Stockholm, Sweden}
\altaffiltext{9}{Department of Physics and Astronomy, Aarhus University, Ny Munkegade 120, 8000 Aarhus C, Denmark}
\altaffiltext{10}{Department of Astronomy, University of Texas at Austin, Austin, TX 78712, USA}

\begin{abstract}

We present imaging polarimetry of the superluminous supernova SN 2015bn, obtained over nine epochs between $-$20 and $+$46 days with the Nordic Optical Telescope.
This was a nearby, 
slowly-evolving Type I  
superluminous supernova  that has been studied extensively and for which  two epochs of spectropolarimetry are also available.
Based on field stars, we determine the interstellar polarisation in the Galaxy to be negligible. 
The polarisation of SN~2015bn shows a statistically significant increase during the last epochs, confirming previous findings.
Our well-sampled imaging polarimetry series allows us to determine that this increase (from $\sim 0.54$\% to $\gtrsim 1.10$\%)  coincides in time with rapid changes that took place in the optical spectrum. 
We conclude that the supernova underwent a `phase transition' at around $+$20 days, when the photospheric emission shifted from an outer layer, dominated by natal C and O, to a more aspherical inner core, dominated by freshly nucleosynthesized material.  
This two-layered model might account for the characteristic appearance and properties of Type I superluminous supernovae. 
\end{abstract}


%%%%%%%%%%%%%%%%%%%%%%%%%%%%%%%%%%  INTRODUCTION   %%%%%%%%%%%%%%%%%%%%%%%%%%%%%%

\section{Introduction}

The study of polarised light is one of the limited tools that allow us to probe the geometry of distant, unresolved supernova (SN) explosions \citep{2008ARA&A..46..433W}.
As photons make their way out of the SN photosphere, they are subject to multiple scatterings that induce polarisation.
If the sky projection of the photosphere is circular, there is no prevalent direction and the SN appears unpolarised.  
However, deviations from spherical symmetry will produce a net, non-zero polarisation signal that can be modelled and predicted \citep[e.g.][]{1991A&A...246..481H,2003ApJ...593..788K,2015MNRAS.450..967B}.
Continuum polarisation is sensitive to the shape of the photosphere while spectropolarimetry can probe the distribution of the different elements in the line forming region \citep[e.g.][]{2009ApJ...705.1139M,2016MNRAS.457..288R}

Almost 10 years after the discovery of Type I  
superluminous supernovae \citep[SLSNe~I;][]{2007ApJ...668L..99Q,2009ApJ...690.1358B,2009Natur.462..624G,2011Natur.474..487Q}, the question of what powers their luminosity remains unsolved, although several models have been proposed  \citep[e.g.][]{2007Natur.450..390W,2010ApJ...717..245K,2012ApJ...746..121C,Metzger15,2016ApJ...829...17S}.
In \cite{2015ApJ...815L..10L} we presented the first polarimetric observations of a SLSN~I: 
we used the Very Large Telescope (VLT) to observe LSQ14mo at $z = 0.256$. 
Our five epochs of imaging polarimetry, obtained between $-7$ and $+19$ days from peak ($V \sim 19.4$ mag), did not show any significant evolution or polarisation that can be safely attributed to the SN. 
Polarisation measurements for such faint targets are challenging, as achieving the required signal-to-noise ratio (S/N) is difficult even using the largest telescopes.
In addition, LSQ14mo was a `fast-evolving' \citep{2012Sci...337..927G,2015MNRAS.452.3869N} SLSN~I that faded quickly \citep[see also][]{2016arXiv161109910C}, thereby increasing this practical difficulty.

SN 2015bn is one of the most nearby ($z = 0.1136$) and well-studied SLSNe~I. \cite{2016ApJ...826...39N} presented a multi-wavelength study  showing it was a slowly-evolving event 
peaking at $M = -22$ mag.
Furthermore, the light curve showed significant structure, deviating from a smooth rise and decay. We adopt here the terms introduced by \cite{2016ApJ...826...39N} to describe the major undulations in the light curve: the `shoulder' is a plateau around $-$20 days, while the `knee' is a prominent bump at $+$45 days.
Early-phase spectroscopy of SN 2015bn showed that the spectrum went through a transformation somewhere between $+$7 and $+$20 days, while all spectra before and after this period showed a very high degree of similarity between them \citep{2016ApJ...826...39N}. The nebular spectra of SN~2015bn were found to be similar to a number of energetic SNe Ic \citep{2017ApJ...835...13J}, leading \cite{2016ApJ...828L..18N} to suggest that the explosion was powered by a central engine, similar to long gamma-ray bursts.
In addition, due to its proximity and intrinsic brightness, SN  2015bn is the first SLSN~I for which it was possible to obtain spectropolarimetry:
 \cite{2016ApJ...831...79I} presented two epochs obtained with VLT  at $-$23.7 and $+$27.5 days.
These authors found that the SN was characterised by a dominant axis and that polarisation increased considerably between the two epochs. 
Their modelling showed that the data are consistent with an axisymmetric geometry. The increase in polarisation was attributed to the photosphere receding into more asymmetric layers.  \cite{2016ApJ...831...79I}   also favour an explosion powered by a central engine.

Here we present nine epochs of imaging polarimetry of SN~2015bn obtained with the Nordic Optical Telescope (NOT). 
By complementing these data with the existing literature datasets of SN~2015bn, we attempt to obtain a better insight on its nature.
The next section describes our observations and Section 3 how the polarisation was determined. 
Our results are discussed in Section 4.

%%%%%%%%%%%%%%%%%%%%%%%%%%%%%%%%%%  OBSERVATIONS   %%%%%%%%%%%%%%%%%%%%%%%%%%%%%%

\section{Observations} \label{sec:obs}

\begin{figure}
%%\epsscale{.80}
\plotone{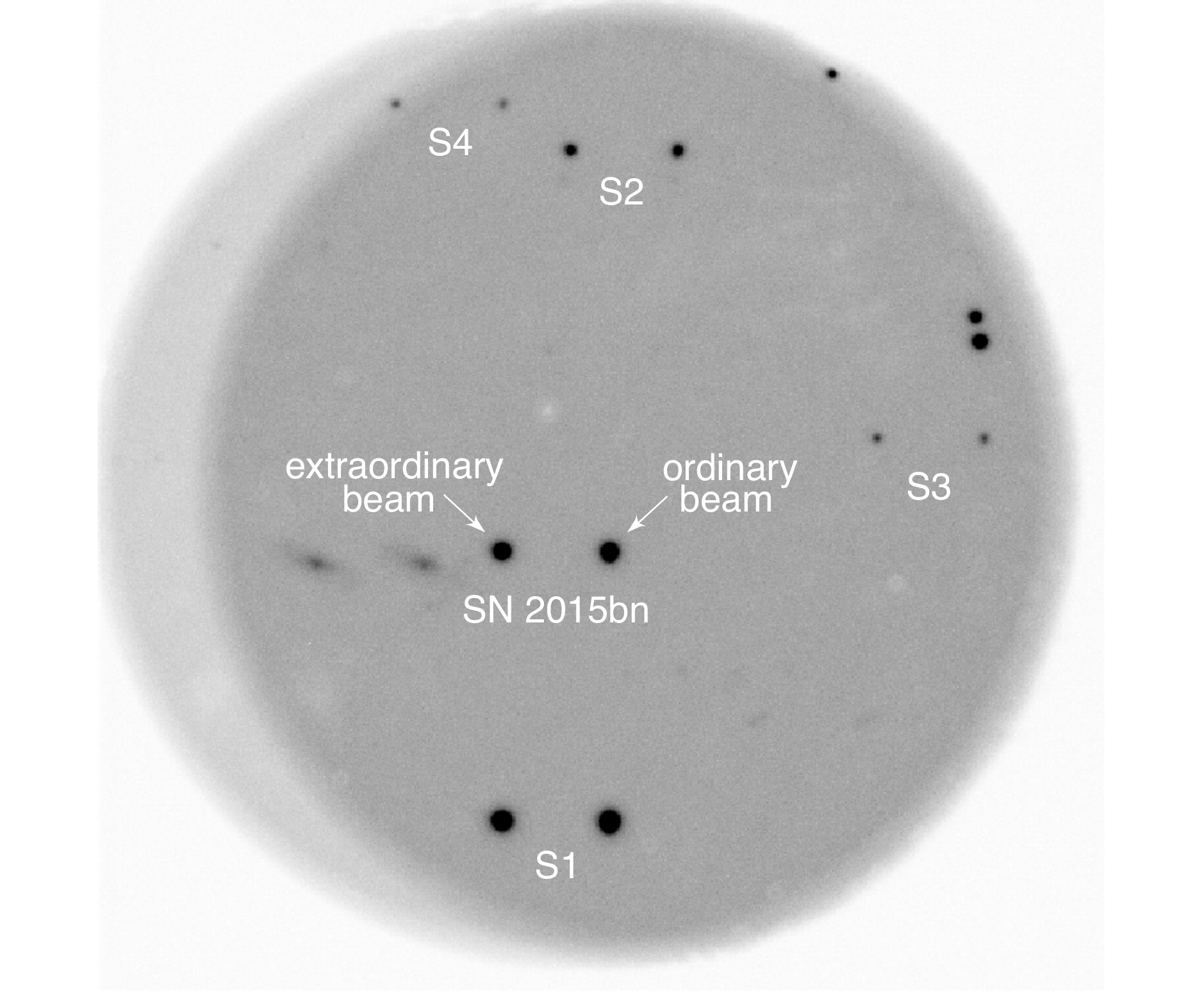}
\caption{SN 2015bn observed with NOT+ALFOSC in polarimetric mode on 16 March 2015, close to maximum light.
Each object appears twice on the image with a separation of 15\arcsec.
There are four more point-like sources (marked S1 - S4) for which both the ordinary and the extraordinary image appear in the FOV.
S1 has an adequate S/N and was used for the determination of the Galactic ISP.
\label{fig:FOV}}
\end{figure}

We used the Andalucia Faint Object Spectrograph and Camera (ALFOSC) instrument at the NOT to observe SN 2015bn in polarimetric mode. 
ALFOSC offers more than one option for linear polarimetry.
Our observations were executed using a half-wave retarder plate in the FAPOL unit and a calcite plate mounted in the aperture wheel, a setting that is suitable for point-like sources, such as our target. 
This setting provides a simultaneous measurement of two orthogonally polarized beams, where the ordinary and extraordinary components appear on the same image, separated by 15\arcsec.  
Figure~\ref{fig:FOV} shows a typical image of SN 2015bn obtained during our observing campaign. 
The field of view (FOV) is approximately circular with a radius of $\sim$1.1\arcmin, and it contains only a limited number of objects other than the SN.
This is both an advantage, as there is no confusion and overlapping between the beams, but also a disadvantage 
because there are few comparison stars that can be used for calibration. 
Specifically, we identify only 4 point-like sources (S1 - S4) whose point-spread functions (PSF) lie fully within the FOV for both beams.
However, S2 and, especially, S3 - S4 are too faint to serve as useful calibrators.

Observations were obtained at four half-wave retarder plate angles ($0$, $22.5$, $45$ and $67.5$ deg) and with the $V$-band filter.
Nine epochs were obtained in total.
A log of our observations is presented in Table~\ref{tab:polalog}. 
Rest-frame phases have been calculated assuming the time of maximum MJD $=$ 57102 \citep{2016ApJ...826...39N}.
The images were reduced in a standard manner, using bias frames and flat field frames obtained without the polarisation units in the light path.

The PSF of the two beams is different as it appears on the detector. In particular, the full width at half maximum (FWHM) as measured on the ordinary beam is typically found to be 1-2 pixels larger than the extraordinary beam in every image obtained. In addition, for a couple of epochs, the PSF did not appear circular but elongated. For this reason, and given the limited number of bright objects in the field, we decided to do aperture photometry in order to measure the fluxes of the SN and the comparison stars (rather than model the PSF of the two beams separately). We used apertures between 2-2.5 $\times$ FWHM of the (ordinary beam) PSF, as we determined that this aperture size was a good compromise in recovering most of the flux, without introducing too much noise in our measurements. We experimented with other aperture sizes and, for the vast majority of images, the results we obtained (location on the $Q$--$U$ plane) were consistent (within the errors) with those obtained for our reference apertures. 

Our procedures were verified by observing the polarimetric standards  BDp64106 and HD204827 and recovering the literature values \citep{1992AJ....104.1563S}.

%%%%%%%%%%%%%%%%%%%%%%%%%%%%%%%%%%  RESULTS   %%%%%%%%%%%%%%%%%%%%%%%%%%%%%%

\section{Determination of the polarisation} \label{sec:disc}

\begin{figure*}
\epsscale{1.27}
\plotone{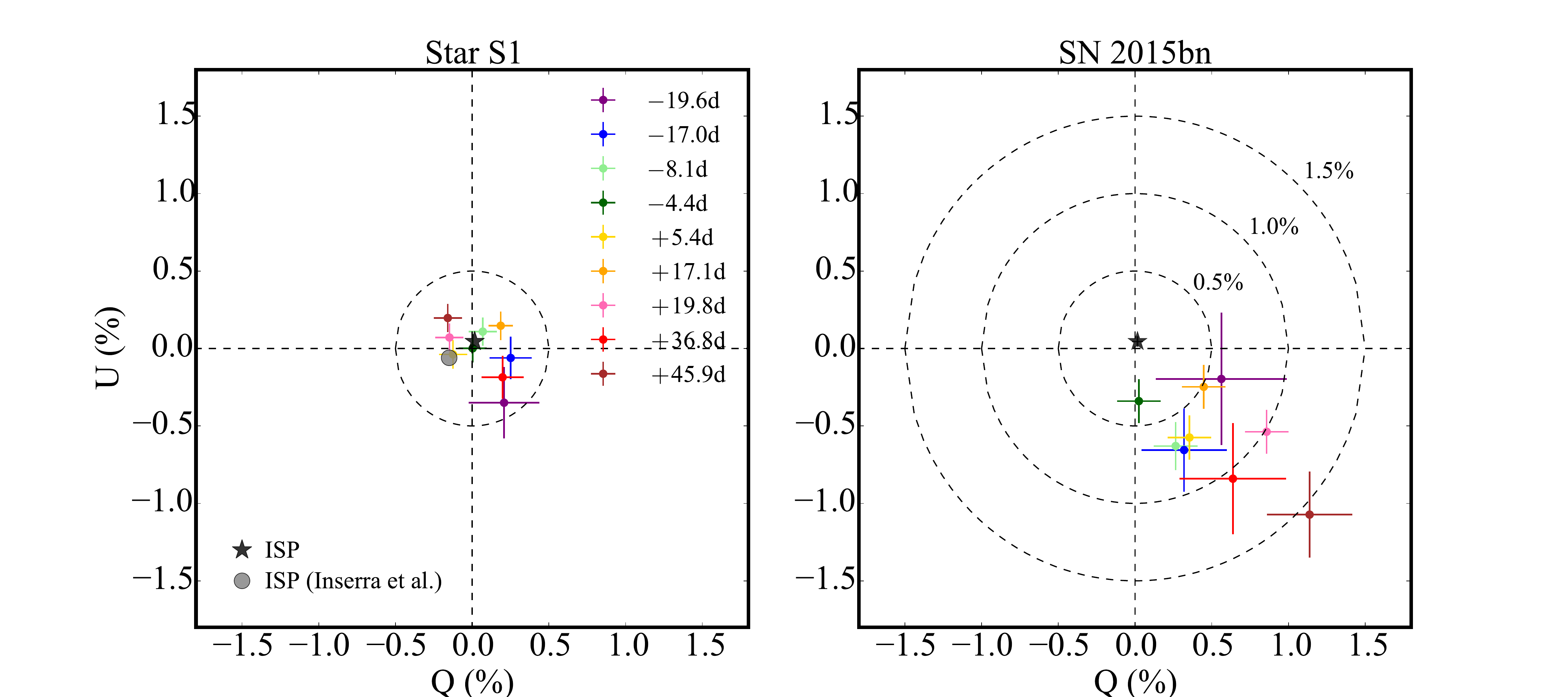}
\caption{The Stokes $Q$--$U$ plane for star S1 (left) and SN 2015bn (right). The different epochs are colour coded as indicated.
The black star shows the location of the ISP as determined by S1. 
For comparison, the ISP estimate by \cite{2016ApJ...831...79I} is also shown. In both cases, the ISP is found negligible.
The SN data on the right panel have been corrected for  the ISP.
Concentric circles of equal polarisation have been drawn to guide the eye.
The last three measurements indicate an increase in the SN polarisation with time.
\label{fig:QUplot}}
\end{figure*}

Based on the normalised flux differences between the ordinary and extraordinary beams, we calculated the Stokes parameters and their errors following \cite{2006PASP..118..146P}. Figure~\ref{fig:QUplot} shows the positions of S1 and SN 2015bn on the Stokes $Q$--$U$ plane, and their evolution.

Stars are generally intrinsically unpolarised and  they can therefore be used to estimate the interstellar polarisation (ISP) towards SN~2015bn. The ISP is not expected to be significant as the Milky-Way extinction towards SN~2015bn \citep{2011ApJ...737..103S} predicts $P_{\rm{ISP}} < 0.2\%$ \citep{1975ApJ...196..261S} and the reddening at the host is found to be negligible, based on the Balmer decrement \citep[][]{2016ApJ...826...39N}, similar to other SLSN~I host galaxies   \citep{2014ApJ...787..138L,Leloudas15,2016ApJ...830...13P}.
In addition, \cite{2016ApJ...831...79I} examined three stars from the \cite{2000AJ....119..923H} catalog within three degrees of the SN position; these stars had polarisation degrees of only 0.07, 0.10 and 0.05\%, respectively.
These authors estimate a $Q_{\rm{ISP}} = -0.15\%$, $U_{\rm{ISP}} = -0.06\%$ directly from their spectropolarimetry of SN~2015bn examining two spectral regions of assumed zero intrinsic polarisation. 
Here, we estimate the Galactic ISP by using star S1 which is bright and shows a statistically non-variable location on the  $Q$--$U$ plane (Fig.~\ref{fig:QUplot}, left).
By taking the weighted average between the different epochs, we obtained $Q_{\rm{ISP}} = 0.02 \pm 0.03\%$ and $U_{\rm{ISP}} = 0.05 \pm 0.03\%$, which we adopt as our best estimate for the Galactic ISP.
Adding more comparison stars does not alter this estimate significantly as their S/N is considerably lower  
and the ISP determination is heavily weighted by S1. 
As explained above, we assume the ISP at the host to be negligible.

\begin{figure*}
%\epsscale{1.27}
\plotone{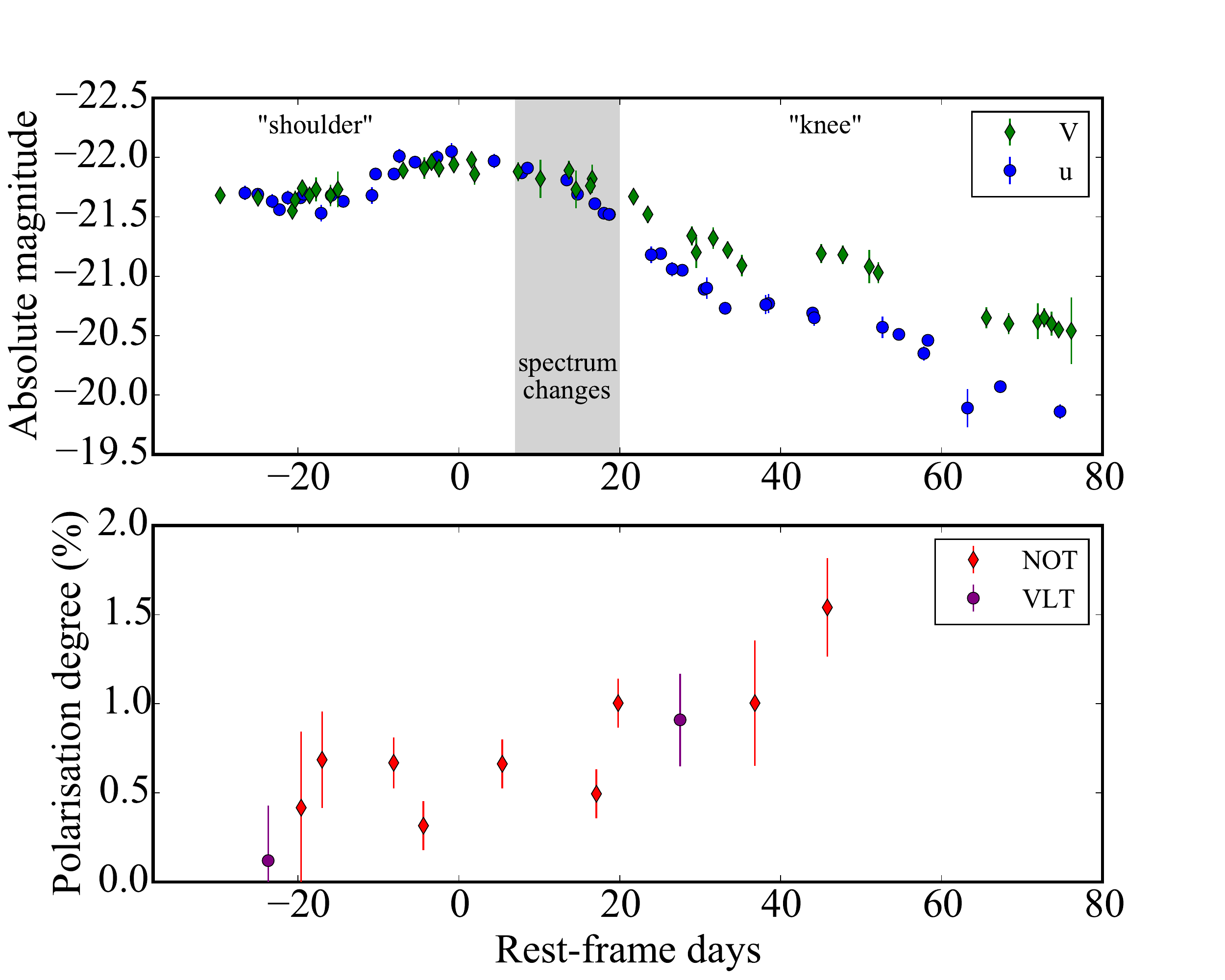}
\caption{Different stages in the evolution of SN~2015bn. In the upper panel, we plot the ultraviolet and visual absolute light curves from \cite{2016ApJ...826...39N} and we mark the two major undulations in the light curve. The shaded region highlights the time interval where changes occurred in the optical spectrum \citep{2016ApJ...826...39N}. 
The lower panel shows the simultaneous evolution of the polarisation degree from our $V$-band imaging polarimetry (red). In addition, we show the integrated $V$-band polarisation from the two epochs of VLT spectropolarimetry \citep[purple;][]{2016ApJ...831...79I}. The data obtained from day $+20$ and onwards show a statistically significant increase in polarisation.
\label{fig:Pevol}}
\end{figure*}

We then corrected the Stokes parameters of SN~2015bn for the (minimal) ISP determined by S1. 
The corrected values are shown in Fig.~\ref{fig:QUplot} (right) and can also be found in Table~\ref{tab:polalog}. 
Finally, from the corrected $Q$ and $U$ we have calculated the polarisation angle $\chi$ and the polarisation degree $P$, including a correction for polarisation bias \citep{2006PASP..118..146P}.
The evolution of the polarisation degree with time is plotted in Fig.~\ref{fig:Pevol}, together with other important milestones in the evolution of SN~2015bn.

%%%%%%%%%%%%%%%%%%%%%%%%%%%%%%%%%%  DISCUSSION   %%%%%%%%%%%%%%%%%%%%%%%%%%%%%%

\section{Discussion} \label{sec:disc}

Both figures~\ref{fig:QUplot} and \ref{fig:Pevol} indicate that polarisation increased with time.
To test this, we performed a $\chi^2$ test for the null hypothesis that $P$ is constant with time. 
This hypothesis can be rejected at a significance $> 3\sigma$  with a $p$-value $= 0.001$.
We then tested for linearity between $P$ and time.
The Pearson correlation coefficient between the two variables is $R =  0.69^{+0.13}_{-0.20}$ but with a $p$-value of $0.04^{+0.14}_{-0.03}$, i.e. the significance of the correlation is somewhere between $1-2\sigma$. 
It is therefore unlikely that the increase in $P$ is linear with time.
We also checked whether the evolution of $P$ can be described by a non-parametric monotonic increase, but a Spearman test revealed that the significance for such a correlation is also low. 
In both tests we have used a Monte Carlo resampling in order to take into account the errors. 
It is instead more likely that polarisation started increasing after a specific point in time. 
The $\chi^2$ test reveals that the polarisation for the first 6 epochs can easily be considered constant, with a $\chi^2/\mathrm{d.o.f.} = 0.93$ around $P = 0.54 \pm 0.07$\%. 
The $p$-value for the hypothesis that $P$ is constant during this period is 0.46.
It is by adding incrementally the last three epochs that the $p$-value reduces to 0.032, 0.037 (i.e. $> 2\sigma$) and, finally, 0.001.
The last data points could be both (visually) consistent with a monotonically increasing trend after day +20, but also with a jump to an elevated polarisation level of $P = 1.10 \pm 0.12$\%.

This increase in polarisation is consistent with the findings by \cite{2016ApJ...831...79I} in their two epochs of spectropolarimetry. 
Of course, these authors have shown that there is a strong wavelength dependence in the polarisation properties of SN~2015bn:
in their first epoch, $P$ increases from $\sim 0.5$\% in the regions below 6300 \AA\ to above $1$\% in the red part of the spectrum.
Their second epoch shows the opposite wavelength dependence where the polarisation is higher ($P \sim 1.4$\%) in the blue but reduces towards the red reaching  $P \sim 0.5$\%.
However, there is an overall increase in the signal that is consistent with what is reported here.
By integrating the polarisation spectra over the NOT $V$-band filter (C. Inserra and M. Bulla, private communication), we obtain $P = 0.12 \pm 0.07$\% and $0.91 \pm 0.07$\% respectively, where we have corrected these values for polarisation bias. 
Adding these points to our polarisation light curve (Fig.~\ref{fig:Pevol}) confirms and strengthens our results.
In addition, \cite{2016ApJ...831...79I} found the existence of a dominant axis on the $Q$--$U$ plane with a position angle that did not change significantly between the two epochs.
Although our observations are obtained in a single wavelength band, our multi-epoch data on the $Q$--$U$ plane also show the clear existence of  a preferred direction that does not change significantly with time (see also Table~\ref{tab:polalog}). 
By fitting all epochs simultaneously, we obtain a linear fit with a slope of $-0.32 \pm 0.29$, corresponding to a position angle of $-17^{\circ} \pm 14^{\circ}$.
We can therefore confirm also this finding by \cite{2016ApJ...831...79I}, supporting an axisymmetric configuration whose orientation does not change significantly with time.

Our dense  time sampling allows us to study in more detail the polarisation signal from SN~2015bn in relation to its overall evolution. 
We first investigate whether there  is any connection between the polarisation and the light curve undulations. 
Figure~\ref{fig:Pevol} shows that the polarisation (and therefore the degree of asphericity) is similar during the pre-maximum `shoulder' and at the peak of the light curve. 
On the other hand, our last datapoint shows the highest polarisation value during the post-maximum knee, although the increase in the signal had already started during the smooth decay. 
If really associated, this jump in continuum polarisation could provide an important clue on the nature of the knee, as what causes it should also explain an increase in the asymmetry of the emitting photosphere. 
Unfortunately, we do not have data past the knee to test this hypothesis. 

A perhaps appealing suggestion is that the increase in polarisation is related to the spectral transformation of SN~2015bn. 
\cite{2016ApJ...826...39N} observed that the spectrum of SN~2015bn was characterised by a very slow evolution both before $+7$ and after $+20$ days. 
However, between these phases, the spectrum underwent a rapid change from being dominated by \ion{Fe}{3} and \ion{O}{2} lines in the blue, to lower ionisation lines of \ion{Fe}{2} and \ion{Mg}{1}]. In the red, lines of [\ion{Ca}{2}], \ion{Ca}{2} and [\ion{O}{1}] appeared, replacing the earlier evidence for \ion{C}{2}. 
Although such changes have been observed before in SLSNe~I \citep[e.g.][]{2010ApJ...724L..16P,2011Natur.474..487Q,2016arXiv161207321L}, \cite{2016ApJ...826...39N} showed that in the case of SN~2015bn this dramatic change occurred over a very short period of time. Here, it has been possible to show that another significant change in the properties of SN~2015bn, the increase of polarisation, took place during the same period (Fig.~\ref{fig:Pevol}). 
While the spectral changes can generally be explained by a decrease in the temperature affecting the ionisation of the ejecta, this effect alone cannot explain the simultaneous increase of asphericity. 
The most natural explanation for the increase in polarisation is that the inner layers of the explosion are more aspherical  \citep[also proposed by][]{2016ApJ...831...79I}.
Such an effect is to some degree expected and has also been seen for other SN types, including SNe~IIP after the plateau phase, when the photosphere recedes inside the external H envelope \citep{2006Natur.440..505L,2010ApJ...713.1363C} and SNe~IIb \citep{2007ApJ...671.1944M}.

The simultaneous rapid spectral and geometric changes suggest that SN~2015bn went through a `phase transition' rather than a smooth evolution. 
We propose here that the ejecta consist of two separate layers with different geometry and possibly different composition: 
an outer layer composed primarily of C and O and natal trace elements and an inner layer with nucleosynthetically processed ejecta (Fe, Mg, Ca).
It is the outer layer that gives SLSNe~I their special appearance.  
A change of opacity drives the photosphere rapidly inside the inner layer, due either to temperature effects or to a jump in density.
Within this scenario, it is possible that when the photosphere moves inside the inner, more aspherical, layer, the outer layer becomes nebular.
This can explain both the early appearance of nebular lines, such as [\ion{O}{1}] and [\ion{Ca}{2}], and their co-existence with photospheric lines originating from the inner ejecta. 
In fact, [\ion{Ca}{2}] appears non-polarised in the $+27.5$ spectrum of \cite{2016ApJ...831...79I} and [\ion{O}{1}] only mildly so, although blending of the latter with \ion{Si}{2} is an issue at these early phases.
The initial profiles of these lines are asymmetric showing a possible deficiency in the red, which progressively disappears with time as they obtain their final single-peaked profile \citep{2016ApJ...826...39N,2016ApJ...828L..18N}.
This could be consistent with them originating in an outer layer surrounding an optically thick core. 
However, this profile evolution could also have a different physical origin as it appears more often in the spectra of SNe Ib/c  \citep[e.g.][]{2014AJ....147...99M}. 
The existence of multiple emitting regions in the ejecta of slowly-evolving SLSNe is also supported by the detection of nebular lines with different line widths in their spectra \citep{2017arXiv170100941I}.

%%%%%%%%%%%%%%%%%%%%%%%%%%%%%%%%%%  CONCLUSION   %%%%%%%%%%%%%%%%%%%%%%%%%%%%%%

\section{Conclusions} \label{sec:conc}

We have presented nine epochs of imaging polarimetry for SN~2015bn.
We have shown that polarisation increased with time, confirming previous findings by \cite{2016ApJ...831...79I}, based on two epochs of spectropolarimetry.
The highest degree of polarisation (asymmetry) was measured during the light curve `knee'. 
However, it is not certain that there is a physical relation, as we did not observe something similar during the other major light curve undulation (the `shoulder').
Furthermore, we have shown that this increase in polarisation occurred almost  simultaneously with the spectral changes in SN~2015bn. 
This geometrical and spectral transition  suggests that a major change took place in the SN ejecta at about $20$ days past maximum.  
We suggest that the photosphere receded from an outer extended layer of natal progenitor composition, dominated by C and O, to an inner, more aspherical layer, of nucleosynthetically processed ejecta. 
It is possible that during this transition, the outer layer became nebular. 
The stellar evolution path resulting in this characteristic for SLSNe~I configuration is still unknown, but it can be related to the special environments in which SLSNe~I are found.
\citep[e.g.][]{2016arXiv161205978S}.

In \cite{2015ApJ...815L..10L} we searched for a similar polarisation evolution in LSQ14mo but we were not able to find it. It is possible that this was due to the more limited time coverage (our observations stopped at day $+19$), to the increased distance (noise) to LSQ14mo, to a special viewing angle, or simply to the fact that LSQ14mo was different. 
LSQ14mo was a fast-evolving event, which was intrinsically fainter than SN~2015bn and did not show any (broad) undulations in the light curve such as SN~2015bn.
This diversity highlights the necessity to extend polarimetric studies of SLSNe to later phases and more SLSN subtypes.

%%%%%%%%%%%%%%%%%%%%%%%%%%%%%%%%%%  THE REST   %%%%%%%%%%%%%%%%%%%%%%%%%%%%%%

\begin{acknowledgements}
 
We are grateful to Cosimo Inserra and Mattia Bulla for valuable comments and for sending us integrated $V$-band polarisation values from their spectropolarimetry.
AG-Y is supported by the EU/FP7 via ERC grant No.~307260, the Quantum Universe I-Core program by the Israeli Committee for Planning and Budgeting and the ISF; and by Kimmel and YeS awards.
DM is partly supported by the Instrument Center for Danish Astrophysics (IDA).
AdUP acknowledges support from a Ram—n y Cajal fellowship and project AYA2014-58381-P.
MS is supported in part by a Sapere Aude Level 2 grant funded by the Danish Agency for Science and Technology and Innovation and by a grant from the Villum foundation. 
The research of JCW is supported in part by NSF AST-1109801.
Based on observations made with the Nordic Optical Telescope, operated by the Nordic Optical Telescope Scientific Association at the Observatorio del Roque de los Muchachos, La Palma, Spain, of the Instituto de Astrof\'isica de Canarias.

\end{acknowledgements}

\bibliographystyle{apj} 
%\bibliography{sn2015bn.bib}

%\clearpage

\begin{deluxetable*}{cccccccccc}
  %\tabletypesize{\scriptsize}
  %\rotate
  \tablecaption{Polarimetry of  SN 2015bn \label{tab:polalog}}
%\tablewidth{\textwidth}  
\tablehead{
    \colhead{UT} & 
    \colhead{MJD} & 
    \colhead{Phase \tablenotemark{a}} &
   \colhead{Exp. time \tablenotemark{b}} &
    \colhead{FWHM} \tablenotemark{c} &
    \colhead{S/N \tablenotemark{d}} & %&
    \colhead{$Q$ \tablenotemark{e}} &
    \colhead{$U$ \tablenotemark{e}} & 
    \colhead{$\chi$}  &
    \colhead{$P$ \tablenotemark{f}} \\
      (yyyy-mm-dd)     &   (days)      &      (days)     &       (s)     &           (pix.)   &           &    (\%)      &       (\%)     &       (deg)     &    (\%)            }
  \startdata
2015-02-27    &   57080.13    &     $-$19.6  &    90    &    8.6   &      166   &   0.56 (0.43) &    $-$0.20 (0.43) &	 $-$9.6 (20.5) &     0.42 (0.43)     \\
2015-03-02    &   57083.02    &     $-$17.0  &   150    &   3.6    &      261   &   0.32 (0.28) &    $-$0.66 (0.27) &	$-$32.0 (10.6) &    0.69 (0.27)     \\
2015-03-11    &   57092.99    &      $-$8.1  &   300    &   5.7    &      493   &   0.26 (0.14) &    $-$0.63 (0.15) &	$-$33.6 ( 6.0) &    0.67 (0.14)     \\
2015-03-16    &   57097.06    &      $-$4.4  &   300    &   5.6    &      512   &   0.02 (0.14) &    $-$0.34 (0.14) &	$-$42.9 (11.6) &    0.32 (0.14)     \\
2015-03-26    &   57107.98    &      $+$5.4  &   400    &   5.4    &      512   &   0.35 (0.14) &    $-$0.57 (0.14) &	$-$29.1 ( 5.9) &    0.66 (0.14)     \\
2015-04-09    &   57121.06    &     $+$17.1  &   400    &  2.9     &      512   &   0.45 (0.14) &    $-$0.25 (0.14) &	$-$14.5 ( 7.7) &    0.50 (0.14)     \\
2015-04-12    &   57124.09    &     $+$19.8  &   500    &  4.1     &      512   &   0.86 (0.14) &    $-$0.54 (0.14) &	$-$16.0 ( 3.9) &    1.00 (0.14)      \\
2015-04-30    &   57142.93    &     $+$36.8  &   600    &   2.9    &      201   &   0.64 (0.35) &    $-$0.84 (0.36) &	$-$26.4 ( 9.6) &    1.00 (0.35)      \\
2015-05-10    &   57152.96    &     $+$45.8  &   600    &   5.8    &      256   &   1.14 (0.28) &    $-$1.07 (0.28) &	$-$21.7 ( 5.1) &    1.54 (0.28)    
  \enddata
\tablenotetext{a}{Rest-frame days with respect to MJD = 57102.}
\tablenotetext{b}{Per half-wave retarder plate angle.}
\tablenotetext{c}{Measured on the extraordinary beam. The pixel scale is 0.21\arcsec/pix.}
\tablenotetext{d}{Average signal to noise ratio for the SN.}
\tablenotetext{e}{Corrected for ISP as determined by star S1: $Q_{\rm{ISP}} = 0.02 \pm 0.03\%$ and $U_{\rm{ISP}} = 0.05 \pm 0.03\%$.}
\tablenotetext{f}{After correcting for polarisation bias \citep{2006PASP..118..146P}.}
\end{deluxetable*}

\end{document}